\documentclass[10pt,twocolumn]{IEEEtran}
\usepackage{multicol}

\usepackage{color}
\usepackage{xcolor}
\usepackage{soul}
\usepackage{amsmath}
\usepackage{graphics}
\usepackage{setspace}
\usepackage{cite}
\usepackage{latexsym}
\usepackage{float}
\usepackage{epsfig}
\usepackage{multirow}
\usepackage{cite,cases,url}
\usepackage{amssymb}
\usepackage{graphicx}
\usepackage{epstopdf}
\usepackage{balance}
\usepackage{subcaption}
\usepackage{enumerate}
\usepackage[shortlabels]{enumitem}
\usepackage{array}
\usepackage{algorithmic}
\usepackage{algorithm}
\usepackage{enumitem}
\usepackage[normalem]{ulem}
\useunder{\uline}{\ul}{}

\newcolumntype{L}{>{\centering\arraybackslash}m{5cm}}
\newcolumntype{K}{>{\centering\arraybackslash}m{6cm}}
\newcolumntype{P}{>{\centering\arraybackslash}m{2.3cm}}
\newcolumntype{M}{>{\raggedright\arraybackslash}m{2cm}}
\newcolumntype{N}{>{\raggedright\arraybackslash}m{2.5cm}}

\begin{document}

\title{Security Threats and Cellular Network Procedures for Unmanned Aircraft Systems}%

\author{\IEEEauthorblockN{Aly Sabri Abdalla and Vuk Marojevic} \\
\IEEEauthorblockA{Dept. Electrical and Computer Engineering, Mississippi State University,
Mississippi State, MS, USA\\
asa298@msstate.edu, vuk.marojevic@msstate.edu}

\vspace{-8mm}
}

\maketitle

\begin{abstract}

This paper discusses cellular network security for unmanned aircraft systems (UASs) 
and 
provides insights into the ongoing Third Generation Partnership Project (3GPP) standardization efforts 
with respect to authentication and authorization, location information privacy, and command and control signaling. 
We introduce the 3GPP reference architecture for network connected UAS and the new network functions as part of the 5G core network, discuss 
introduce the three security 
contexts, potential threats, 
and the 3GPP procedures. 
The paper identifies research opportunities for UAS communications security 
and recommends critical security features and processes to be considered 
for 
standardization. 

\end{abstract}

\IEEEpeerreviewmaketitle
\section{Introduction}
\label{sec:intro}

The unmanned aircraft system (UAS) technology development and 
market penetration has 
led to 
research and development on cellular connected unmanned aerial vehicles (UAVs). 
UAVs are considered as potential cellular network users for receiving command and control (C2) and other services. 
They may also provide network support to extend coverage, increase capacity, or 
enhanced security in 4G, 5G, and future 6G networks \cite{UAV_WirelessMag}. 

Different standards groups, including The IEEE, The International Telecommunication Union (ITU), and The Third Generation Partnership Project (3GPP), have initiated 
working groups (WGs) to enable the integration of UAVs into cellular networks. 
An recent IEEE WG focuses on developing the 
architecture and protocols for facilitating self-organizing, spectrum-agile communications for UAVs 
to enhance terrestrial connectivity~\cite{IEEE}. An ITU study group standardizes a functional architecture of IMT-2020 networks where both the UAV and its controller (UAV-C), which comprise a UAS, are considered user equipment (UEs)~\cite{R3}. 
The 3GPP has established WGs to identify a reference architecture and the requirements/assumptions for remote identification (RID) and tracking (RID\&T) of UAVs and C2 signaling, 
among others. 
The application layer architecture 
is another 3GPP study item to support efficient UAS deployment and service provisioning. 
The 3GPP has identified a number of connectivity and interference issues for cellular connected UAVs and has recommended 
solutions in its technical reports and standards for 4G and 5G~\cite{3GPP_Stnds}.

Security is vital for 
efficient cellular connected UAV deployments. This includes the confidentiality protection of 
identifiers (IDs), spoofing immunity, and various levels for the integrity 
and privacy preservation of UAS control and data links
~\cite{UAS_5G}. 
The threat model has shifted since sophisticated software radio hardware and software became widely available.  
Targeted wireless attacks to cellular networks, such as eavesdropping, jamming, 
and spoofing of control and data channels, can be implemented with open-source software investments~\cite{MarVTC17,SDRAttack}. The 3GPP has therefore initiated a study on security aspects of network connected UAVs to identify 
key issues and solutions~\cite{UAS_Sec}.

Early research has 
studied the privacy and confidentiality concerns of network connected UASs. 
Alladi et al.~\cite{MAP} propose a physically unclonable function scheme for the lightweight mutual authentication between UAVs and the 5G base station (BS) with unique and secure session keys for each session. Bansal et al.~\cite{AuthK} 
develop a one-to-one (UAV-to-BS) scalable authentication protocol using K-means clustering. 
Li et al.~\cite{Identity} present an elliptic curve cryptography authentication scheme to preserve the ID and authenticate 
the UAV and the ground BS with low computational cost. 

This paper discusses cellular network security for UASs in the broader sense 
and 
provides insights into the ongoing 3GPP standardization efforts 
with respect to authentication and authorization (A\&A), location information privacy, and 
C2 signaling. 
The objective 
is to comprehend 
the reasons behind the 3GPP procedures for securing the UAV and UAV-C network connections and services. 
We therefore explore the critical security contexts and 
potential threats 
before outlining the 3GPP security procedures and 
identifying the remaining research 
and standardization opportunities.

The rest of the paper is organized as follows: Section II introduces the 3GPP reference architecture for network connected UASs and the new 5G network functions (NFs). 
We present the security context and identify the main 
threats and the corresponding 
solutions developed by The 3GPP in Sections III and IV. Section V identifies 
the remaining 
challenges and opportunities for 
research, development, and standardization. Section VI provides the concluding remarks. 

\section{The 3GPP Network 
Architecture for UASs}
\label{sec:requirements}

\begin{figure*}[t]
    \centering
    \includegraphics[width=0.9\textwidth]{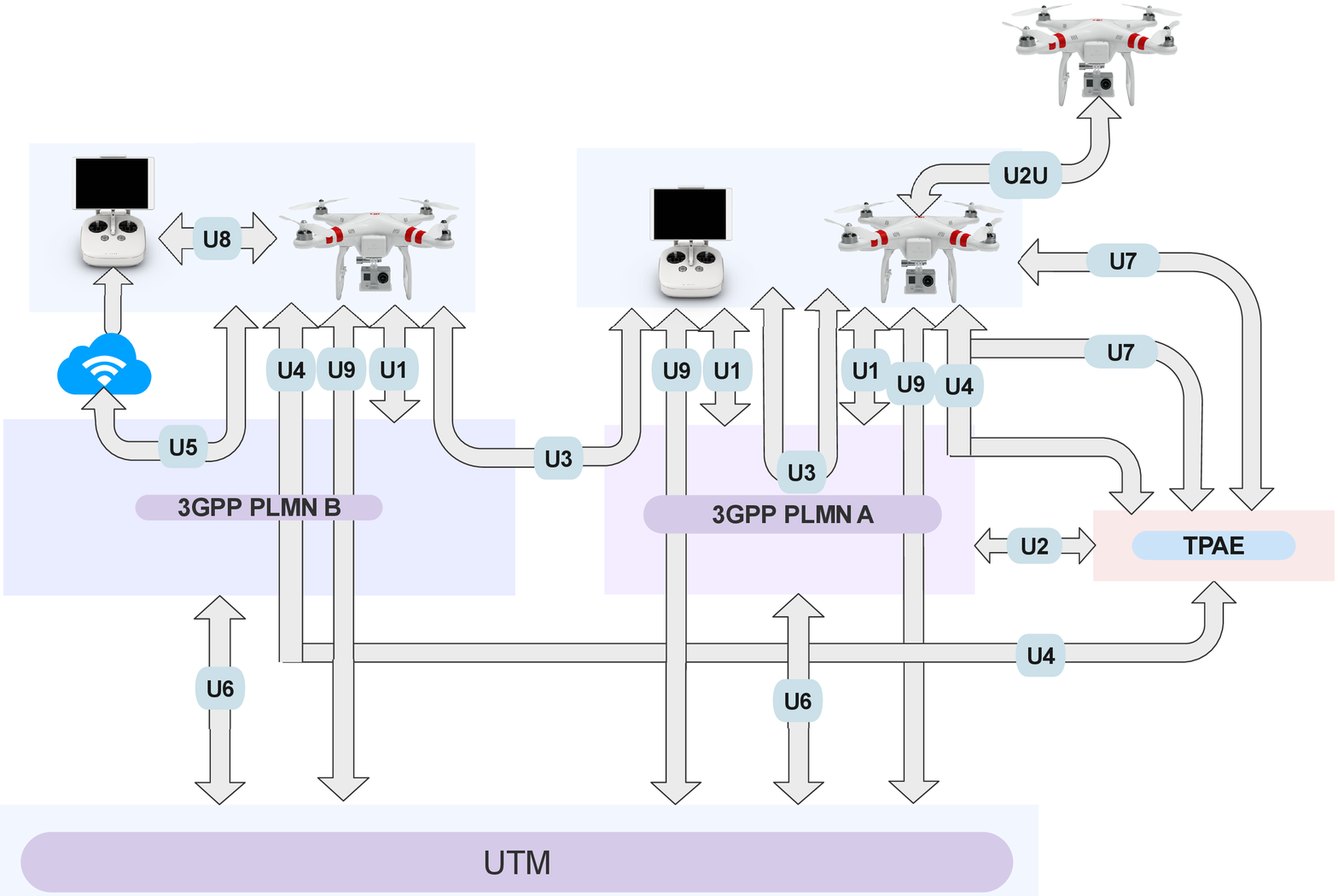}
    \caption{
    The 3GPP reference architecture for cellular network connected UAS (PLMN--public land mobile network; TPAE--third party authorized entity; UTM--UAS traffic management).
    }
    \label{fig:Figure11}
    \vspace{-5mm}
\end{figure*}
This section introduces the reference architecture of the 3GPP standardization body for supporting 
UAS applications and use cases. 

\subsection{
The 3GPP Architecture and Interfaces for UAS Operations}
The 3GPP standardization considers a UAS as a UAV and UAV-C pair, where each will be authorized as an individual UE in the 3GPP network. The 3GPP work items aim to provide a network 
architecture that enables control plane (CP) and user plane (UP) communications services for UASs and provide wireless connectivity between the UAS and non-3GPP aviation entities, such as the UAS Service Supplier (USS) and the UAS Traffic Management (UTM) for beyond visual line of sighs (BVLOS) operations. The USS and UTM entities are responsible for providing various functions to ensure the safety and security of the UAS operations. These functions include C2 services, services to civil aviation authority (CAA), telematics, UAS-generated data, RID, authorization, enforcement, and regulation of UAS operation. The UTM/USS can be integrated in the 3GPP framework as an application function (AF), operating as a CP network function, or an application server in the data network.

For the support and assistance of 
UAS operations, The 3GPP 
is working on 
an architecture that is applicable to the evolved packet system and the 5th generation system. The proposed architecture should be able to associate and identify the UAV-C, whether or not it uses the 3GPP 
network to connect to the UAV. 
The 3GPP architecture for 
UASs provisions interworking between the UAV and UAV-C even if both nodes are served by two different public land mobile networks (PLMNs). 

Figure~\ref{fig:Figure11} illustrates the reference architecture for cellular connected UASs as proposed by the 3GPP standardization group in the 
Technical Report (TR) 23.754~\cite{UAS_5G_identification}. It assumes that there are other external entities that are not included as UTM functionalities and that can monitor UAVs, track UAV data, and control UAVs. These entities are defined under the umbrella of the third party authorized entity (TPAE). The TPAE can be an application server in the data network from the perspective of the 3GPP network. The control of a UAV for BVLOS operations can be performed either by the UAV-C or by the TPAE, where the C2 packets may be exchanged between the UAV and the UAV-C, UTM, or TPAE. 
The 
network interfaces are also illustrated in Fig. \ref{fig:Figure11} and introduced in continuation.

\begin{enumerate}[start=1,label={\bfseries U\arabic*:}]
    \item The interface between the UAV and UAV-C with the 3GPP network for facilitating authorization, authentication, identification, and tracking of the UAV and UAV-C,
    \item The interface between the 3GPP network and the TPAE for facilitating the RID\&T of the UAV,
    \item The interface responsible for transporting C2 packets between the UAV-C and the UAV via intra or inter-PLMN UP connectivity, 
    \item The interface between the UAV node and the TPAE through the 3GPP network for C2 signaling and RID\&T of the UAV, 
    \item The interface 
    for transporting C2 packets between the UAV and the UAV-C, the latter being connected to a non-
    3GPP network via the Internet,
    \item The interface between the 3GPP network and an external USS/UTM entity for enabling identification, authorization, and tracking of the UAV,
    \item The interface between the UAV and other entities outside the scope of The 3GPP for broadcasting the RID, 
    \item The interface between the UAV and the UAV-C for transporting C2 data over a network that is beyond the scope of The 3GPP, 
    \item The direct interface between different components of the UAS (UAV, UAV-C) and the USS/UTM for various operational functions such as networked RID, 
    C2, UAV authentication, authorization, and tracking, and 
    \end{enumerate}
    \begin{enumerate}[label={\bfseries U2U:}]
    \item The interface between two UAVs 
    to support the broadcast of the RID. 
\end{enumerate}
\begin{figure*}[t]
    \centering
    \includegraphics[width=0.95\textwidth]{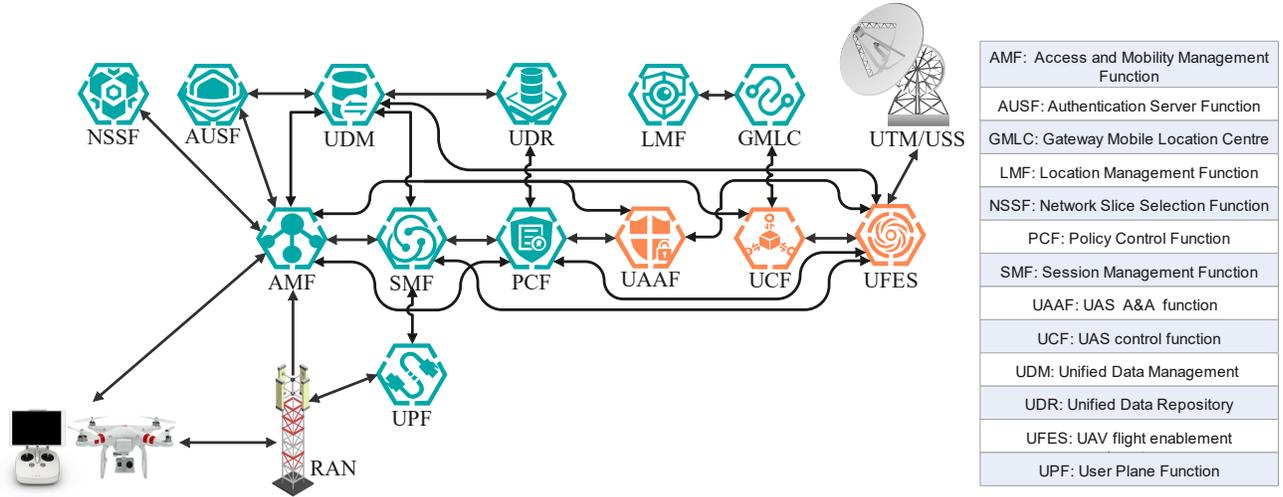}
    \caption{
    The 5GC integrating 
    three UAS network functions. 
    }
    \label{fig:Figure2}
    \vspace{-5mm}
\end{figure*} 
\subsection{
UAS Operations Related 5G Core Network Functions}
For the complete 3GPP network support 
of UAS applications, 
enhancements to the 5G core network (5GC) are necessary. These 
come in the shape of AFs and The 3GPP defines three of them specifically for supporting UAS operations. These are highlighted in Fig.~\ref{fig:Figure2} and introduced in continuation.
\begin{itemize}[leftmargin=+9.4pt]

    \item \textbf{UAV flight enablement subsystem (UFES):} The UFES is implemented to serve as a single interface to the USS/UTM. 
    Principally, the UFES performs the USS/UTM discovery mechanisms and the selection procedures without requiring other 3GPP network nodes. 
    The USS/UTM selection by the UFES is based on the CAA-Level UAV ID, which provides RID\&T information to the TPAE/USS/UTM that may be monitoring a UAV. 
    The UFES supports delivery of the UAV external ID as the 3GPP UAV ID to the USS/UTM, and can retrieve relevant subscription information from the unified data management (UDM) and/or receive policy control information from the policy control function (PCF). The UFES determines a protocol data unit (PDU) session for the UAV operation through the session management function (SMF) to transmit the operation updates from the USS containing the updated authorized UAV and UAV-C pairing information. 
    
     \item \textbf{UAS A\&A function (UAAF):} The UAAF is a new 3GPP AF that assists with the A\&A of UAS nodes over the UP. A UAV originating A\&A request is transferred to the UAAF through the access and mobility management function (AMF). It includes the UAV ID, the UAV application ID, and the served UTM/USS ID.
     The UAAF validates that the UAV has a valid subscription and includes relevant subscription and application information from the PCF to be sent to the UTM/USS via the UFES. 
     
     \item  \textbf{UAS control function (UCF):} The UCF is 
     operated by the PLMN serving the UAV/UAV-C. It is able to deliver the UAV/UAV-C location 
     reports or deferred reports upon request from the UTM/USS. The UCF invokes the gateway mobile location centre (GMLC) procedures for obtaining the location of the UAV or the UAV-C upon receiving a request from the UTM/USS via the UFES, while triggering the AMF for registration information related to the 
     served node. 
     The UCF is also responsible for matching the 3GPP UAV ID provided by the UTM/USS with the 3GPP UE ID and transferring the CAA-level UAV ID to the UTM/USS, where both IDs are needed for successful authentication and location procedures. The UCF determines the needed 5GC NF 
     to be invoked when the interworking between the 5GC and the UTM/USS is needed. 

\end{itemize}
\vspace{5mm}
\section{
Security Contexts 
and 
Potential Threats} 
\label{sec:contribution}
\subsection{UAS A\&A} 
\textbf{Security context:} 
The 3GPP entities such as the gNodeB 
and the AMF must be able to identify the UAV and UAV-C and distinguish its access from other UEs. 
Based on the architectural requirements~\cite{UAS_5G_identification}, 
there are two types of IDs defined for a particular UAV node. 
The designated CAA level UAV ID, which is assigned by the USS/UTM, is employed for RID\&T. 
The 3GPP UAV ID is used for recognizing the UAV; it provides the necessary credentials for the UAV to become an authorized UE and gain access to 3GPP services. 
The core network is responsible for matching 
the CAA-level UAV ID 
to the 3GPP UAV ID. 

An additional factor that must be taken into consideration to preserve a fully authenticated and authorized process for the USS system is the pairing between UAV and UAV-C that takes place at the USS/UTM. The result of this pairing process must be communicated to the 3GPP network. 

UAS authentication and authorization is the prerequisite 
\textcolor{black}{to enable} 
overruling \textcolor{black}{the UAV-C in case of suspicious access after tracking the UAV data} 
\textcolor{black}{by the 
TPAE that can take over the control of the UAV.} 
Consequently, the connection request must be authenticated and authorized by the 3GPP network 
differently from a normal UAV-C, UAV, or UE. The 3GPP network must follow certain policies regarding the unsuccessful authentication and authorization \textcolor{black}{where the UAAF may inform the SMF} to 
prevent the 
registration 
and/or the cancellation of 
\textcolor{black}{illegitimate} PDU sessions \textcolor{black}{by an unapproved UAV or UAV-C}. 

\textbf{Potential threats:} 
A weak UAS authentication process can grant access to an {untrusted} {UAV or UAV-C} 
to receive UAS services via the 3GPP network. This can cause leakage of critical data such as 
UAS system capabilities, location, and encryption keys. 
Unauthorized UAVs may 
attempt to imitate the behavior of legitimate UAVs to launch man-in-the-middle or replay attacks~\cite{BlockchainUAV}. \textcolor{black}{An unauthorized node that is able to 
obtain the 
credentials of authorized nodes could then inject 
false data. 
} 
In a surveillance scenario, for example, an unauthorized UAV may deliberately alter and provide false 
data (e.g. altered pictures or video streams). 

A fake USS/UTM may inject 
messages to the UAS nodes that affect UAV flight operations with the possibility of UAV hijacking. 

A malicious radio node may 
continuously jam 
the communications channels 
to cause bandwidth saturation, hinder the A\&A process of legitimate UAS nodes requesting network access, or cause denial of service of already authenticated nodes. 

    \subsection{Location Information Veracity and Location Tracking Authorization}
    \textbf{Security context:} 
    The UAV is required 
    to notify the USS/UTM entities of its location using one of several 
    forms of location information, including 
    the absolute position, e.g., global navigation satellite system (GNSS) 
    coordinates, and the relative position, such as cell ID or tracking area based coordinates. 
    The reported location information 
    may be used by the USS/UTM to define the optimal set of actions needed to ensure 
    safe 
    aerial operations. 
    The reporting of location information 
    can be 
    verified 
    using 
    UAS application layer mechanisms such as the networked RID. 
    In addition, it is preferable to advocate the position reporting for both UAV/UAV-C and USS/UTM via network assisted positioning mechanisms offered by the 3GPP network. The 3GPP network forwards the estimated location information to the USS/UTM as supplementary data when it is requested. 
    
    There are already various location services that can be used by 
    the UAV or UAV-C in the evolved terrestrial radio access network 
    or next generation RAN (NG-RAN). These include the network-assisted GNSS, 
    downlink positioning, enhanced cell ID, 
    terrestrial beacon system, reference signal time difference, 
    and observed time difference of arrival. 

    \begin{figure*}[t]
    \centering
    \includegraphics[width=0.95\textwidth]{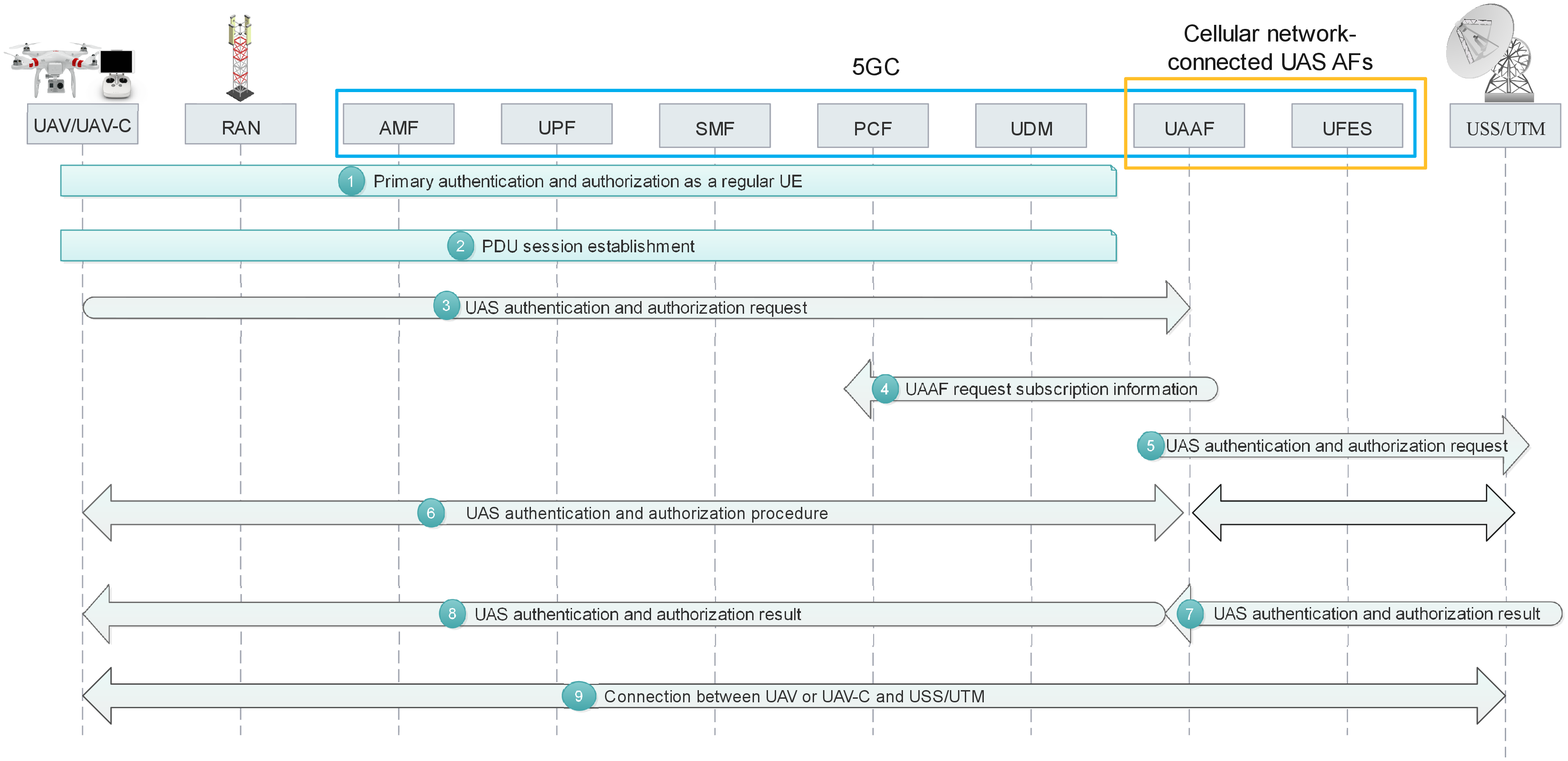}
    \caption{
    The 3GPP workflow for 
    UAV/UAV-C A\&A. 
    }
    \label{fig:Figure4}
    \vspace{-5mm}
\end{figure*} 
    \textbf{Potential threats:} The 
    location information 
    can be compromised through a spoofing attack to create false location reports and force the USS/UTM to mislead the airspace management decisions into inaccurate and dangerous directions. The falsified location data created by a spoofer can lead to costly cyberphysical or kinetic attacks on the 
    UAS and, for example, steer the UAV to fly over unauthorized or prohibited airspace, deceive the maneuver strategy to create air conflicts, or confuse authorities or pilots about the  
    location of UAVs. 
    The location spoofing attacks can be carried out 
    by external means through a fake GNSS or cell ID transmitter
    ~\cite{GPSSPoofing}.
    \subsection{C2 Signaling Integrity}
    \textbf{Security context:}
    The C2 signaling is used to control the UAV operations through the controller, which can be the UAV-C, TPAE, or USS/UTM. The C2 links communications can be provided 
    through the 3GPP 
    network via one of these interfaces---U3, U4, or U9 (Fig.~\ref{fig:Figure11})---as a function of the node that control the UAV. 
    The C2 communications between the UAV and UAV-C can be classified into three modes
    : 
    direct, 
    network assisted, 
    and UTM navigated~\cite{3GPP_Stnds}. 
    It is critical to preserve reliable and available C2 communications in spite of radio condition variations, different traffic situations, and unpredictable events. 
    This must be addressed by means of selecting or  switching to the appropriate C2 communications mode. For example, when a UAV approaches the BVLOS region from the controller and the direct communications link between the UAV and its controller becomes unstable, it is preferable to switch from direct 
    to network assisted C2.   
        
    \textbf{Potential threats:} The ability to eavesdrop, monitor, or otherwise attack 
    C2 communications between the UAS peers is a 
    security risk that must be suppressed to ensure the safety and integrity 
    of aerial operations.  Uncertainty in the security measures for C2 links makes the system vulnerable to control deficiencies that can lead to failures of operation or UAVs being hijacked. 
    Smart attackers can target and take advantage of the 
    switching process 
    between the C2 modes 
    and exploit the security vulnerabilities of the least protected mode. 
    A combined eavesdropping and jamming attack can be conducted over the C2 links, where the jammer downgrades the QoS and initiates the process of switching from one C2 mode to another. 
    The eavesdropper may then 
    intercept the 
    control messages and use this information to further attack the system. 


\section{3GPP Security Solutions}
\label{sec:Challenges}

This section introduces the 3GPP approaches that were designed to prevent many of the previously described 
security threats. Specifically, we discuss the 3GPP procedures to secure access, location information, and C2 signaling.

\subsection{UAS A\&A}
Figure~\ref{fig:Figure4} presents the workflow suggested by The 3GPP for 
the UAS A\&A. 
It involves the UAAF, which is a new AF that is used to validate the subscription information of the UAV and UAV-C and assist with the A\&A process of the USS/UTM. The procedures is described in continuation:

\begin{enumerate}
    \item The primary A\&A is performed between the UAV/UAV-C and the 5G network just like a regular UE does through the PLMN UE ID (i.e., the subscription permanent ID) 
    and the corresponding credentials.  
    \item A PDU session is established between the UAV/UAV-C and the UAAF for enabling the 
    UAS specific A\&A message exchanges with a default policy that prevents any traffic from the UAV/UAV-C except the traffic destined for the UAAF.
    \item The UAV/UAV-C initiates the A\&A request with the UAAF as UP data providing 
    the UAV/UAV-C identity, USS/UTM identity if already known, and application level information. 
    \item The UAAF request the relevant subscription information of the UAV/UAV-C node that initiated the A\&A process from the PCF with the assistance from the 
    binding support function (BSF), which 
    binds the UAV/UAV-C application function request to the PCF. 
    \item  After receiving the subscription information of the UAV/UAV-C from the PCF, the UAAF checks its validity for aerial subscription and, if the check is successful, 
    the UAAF determines the USS/UTM serving the UAV/UAV-C based on the provided information in Step 3 and the 
    list stored in the UAAF with valid USS/UTM identities. The 3GPP UAV ID that is obtained from the BSF 
    is then added to the CAA-Level UAV-ID and the application level information and forwarded to the USS/UTM. This completes 
    the A\&A process. The UFES facilitates the communications between the UAAF and the USS/UTM.
    \item If the information sent to the USS/UTM 
    in the previous step is not sufficient for 
    A\&A, 
    the UAAF relays the needed messages between the UAV/UAV-C and USS/UTM through the UFES 
    to the UAV application and USS/UTM.
    \item 
    The A\&A result becomes transparent and is provided to the UAAF. 
    If the authentication is successful, the USS/UTM may provide application specific information to be used for secure communications. 
    If the authentication is unsuccessful, the USS/UTM may inform the UAAF about the possible measures to take, e.g. to terminate the PDU session established in Step 2.
    \item The UAAF relays the 
    result 
    to the UAV/UAV-C through the 5GC. 
    \item If the result of Step 6 is successful, the UAAF informs the SMF to modify the PDU session established in Step 2 {with the authenticated identities of the UAV/UAV-C} such that the UAV/UAV-C can communicate with the USS/UTM, that is, beyond the limitations of the prior PDU session mentioned in Step 2. 
\end{enumerate}
    \begin{figure*}[t]
    \centering
    \includegraphics[width=0.95\textwidth]{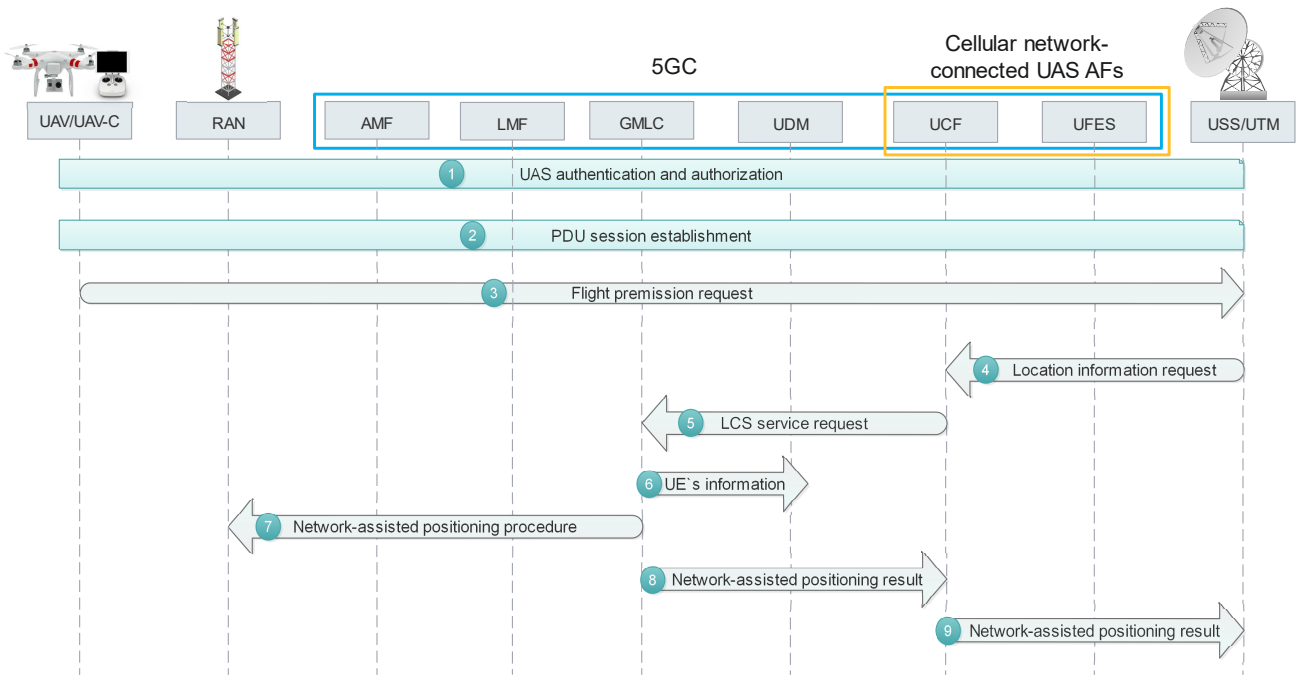}
    \caption{
    The 3GPP workflow 
    for UAV/UAV-C location information provisioning, verification, 
    and tracking. 
    }
    \label{fig:Figure5}
    \vspace{-5mm}
\end{figure*} 
\subsection{Location Information Veracity and Location Tracking Authorization}
Figure~\ref{fig:Figure5} presents the 3GPP workflow for the secure exchange of location information. 
This workflow involves 
the UCF which is responsible for the location verification and for tracing the information of the UAV and UAV-C to provide trustful location reporting to the USS/UTM. The workflow is described below:

\begin{enumerate}
    \item The process starts with the primary A\&A process of the UAS node as a UE in the 5G network followed by the A\&A 
    with the USS/UTM 
    to validate the aerial subscription as described in the previous subsection and illustrated in Fig. \ref{fig:Figure4}. 
    \item The 5G system will establish the PDU session 
    for location information and tracking data exchange and validation 
    between the UAV/UAV-C and the USS/UTM. 
    \item The UAS node sends the flight operation permission request as UP data to the UTM. 
    This request may include the UAV identity, its current location, planned trajectory, and so forth.
    \item The USS/UTM initiates the location request and verification procedures by communicating with the UCF through the UFES. 
    The location information request includes the \textcolor{black}{CAA-level UAV ID}. 
    \item After receiving the request, the UCF activates the location services AFs of the 5GC through the GMLC to trigger the location verification procedures as requested by the USS/UTM in the previous step 
    and to obtain the location information of the UAV and the corresponding UAV-C by following the location procedures defined in 3GPP TS 23.273. 
    \item The GMLC then invokes a service operation request in the UDM for the target node to obtain the privacy settings. 
    The UDM returns the network address of the 
    serving AMF. 
    \item The GMLC communicates with the location management function (LMF) to select the network assisted positioning method which relies on the location measurement from the NG-RAN nodes, i.e. NG-BSs. The LMF invokes the service operation towards the AMF to request the transfer of a network positioning message to a NG-BS. 
    The target NG-BS obtains and returns the position related information. Then the LMF calculates the location result and responds to the GMLC. 
    \item The obtained 
    \textcolor{black}{location measurement} is transferred from the GMLC to the UCF. 
    \item The UCF finally forwards the 
    \textcolor{black}{
    location measurement} 
    obtained in Step 7 to the USS/UTM through the 
    UFES. This information can be used to verify the location or flight behavior that the UAV reported. 
\end{enumerate}
\begin{figure*}[t]
    \centering
    \includegraphics[width=0.95\textwidth]{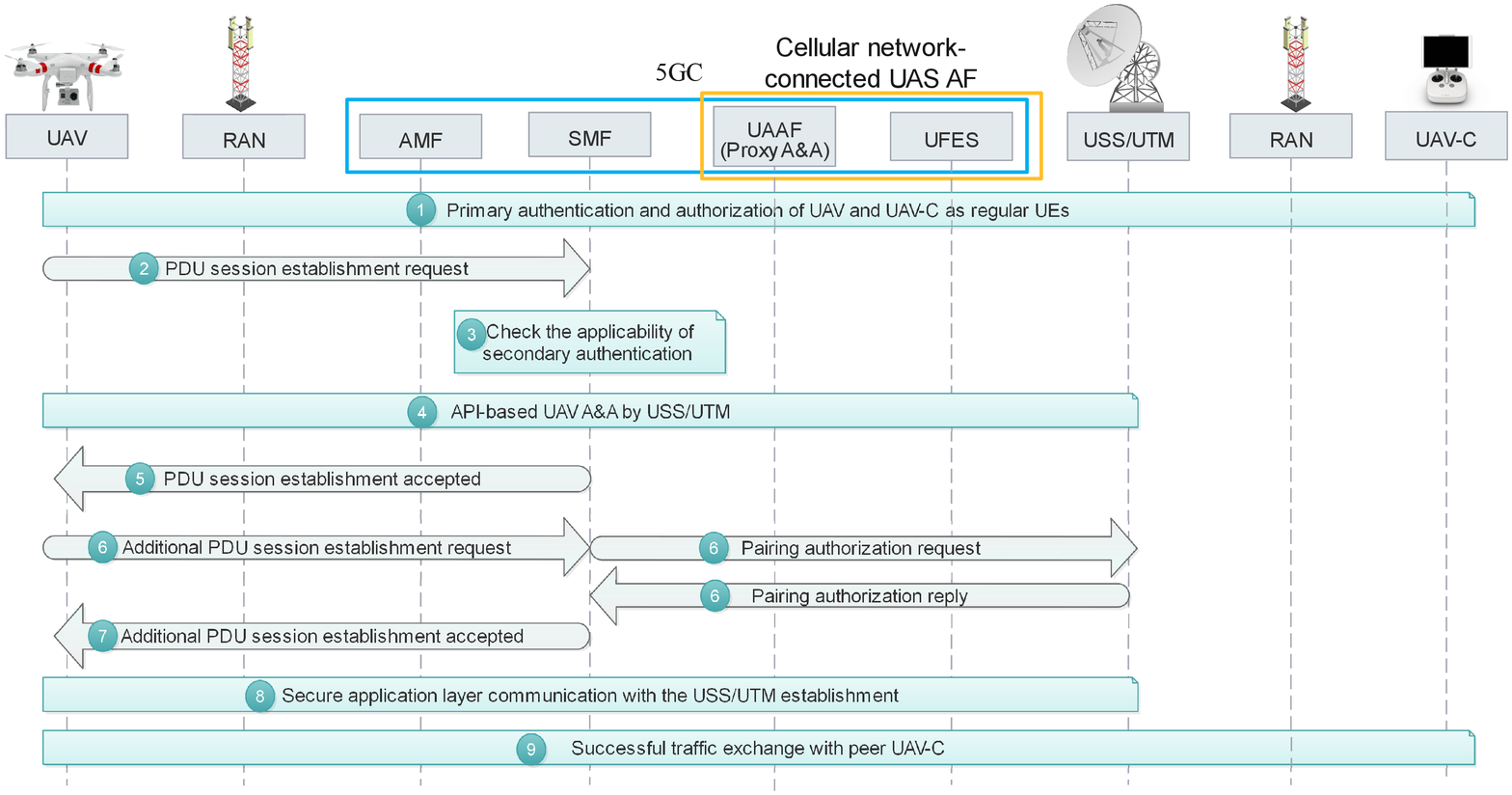}
    \caption{
    The 3GPP workflow 
    to authenticate and authorize C2 communications between UAV and UAV-C over the 5G network. 
    }
    \label{fig:Figure6}
    \vspace{-5mm}
\end{figure*} 
\subsection{C2 Signaling Integrity}
Figure~\ref{fig:Figure6} illustrates the 3GPP workflow 
for secure 
C2 communications link establishment between the UAV 
and the its controller, which can be the UAV-C or the USS/UTM.  
This procedure is an application programming interface (API)-based solution that facilitates a secondary authentication with the USS/UTM for the PDU sessions via the 5G data network authentication, authorization and accounting 
server. The following steps are performed:
\begin{enumerate}
    \item The primary authentication procedure is performed between the UAV and the 5G network and between the UAV-C and the 5G network 
    for registering with the network as regular UEs.
    \item A request message is sent from the UAV to the AMF for establishing the PDU 
    session with the USS/UTM. This message includes the CAA-level UAV ID and data network name/single-network slice selection assistance information (DNN/S-NSSAI). The AMF uses the subscription information of the UAV and the DNN/S-NSSAI to determine the appropriate SMF.
    \item The SMF performs a check on the applicability of the requesting UAV 
    to perform the secondary authentication based on the supplied subscription information and the local policies. 
    \item The SMF 
    triggers the USS/UTM to initiate the API-based authentication process through a proxy A\&A function implemented by the UAAF AF in the 5GC. The USS/UTM address can be resolved by the SMF with the obtained CAA-level UAV ID. 
    Next, the proxy A\&A initiates communications 
    with the USS/UTM through the UFES 
    for performing the secondary authentication and sends the 3GPP UAV ID to the USS/UTM.  If the process succeeds, the USS/UTM sends back a new assigned CAA-level UAV ID and authorization token and/or key material to the proxy A\&A. These new credentials are provided by the proxy A\&A to the SMF. 
    \item The SMF sends the PDU establishment session accept message to the UAV with the new credentials. 
    \item An additional PDU session establishment request 
    is initiated by the UAV 
    by sending the request to the SMF. The new PDU session will be enabled by providing the new 
    CAA-level UAV ID obtained from the secondary authentication procedure and the UAV-C identity. The PDU session includes a pairing authorization request with the UAV-C to the USS/UTM. The USS/UTM informs the SMF with the authorized Internet protocol address of the UAV-C that has been 
    obtained by the pairing authorization process 
    to reconfigure the PDU session accordingly. 
    \item The UAS communications is confirmed by sending 
    the PDU establishment accept message to the UAV, which will apply the new credentials and security parameters 
    for future communications.
    \item The secure application layer communication is established between the UAV and the USS/UTM 
    for C2 
    communications while using the new security parameters (authorization token and/or key material).  
    \item The UAV initiates secure C2 communications with the peer UAV-C.
\end{enumerate}

\section{
Remaining Challenges and Future Directions 
}
\label{sec:Challenges}
We have reviewed the 
standardization efforts for facilitating secure 
cellular connected UAS contexts. Various challenges remain 
that 
we identify in continuation 
and that we recommend to be considered as research and standardization work items. 
\begin{itemize}[leftmargin=+9.4pt]
    \item \textbf{
    Encryption:} The broadcast nature of communications links between the UAV and the UAV-C or among 3GPP entities for both payload data and C2 packets, make them vulnerable to eavesdropping and adversarial attacks. Encryption of transmitted signals among UAS entities has not been standardized yet within the scope of The 3GPP, but there 
    are efforts 
    that address this problem as part of open source and commercial software projects for UAVs. For example, the Paparazzi and DJI open source UAV projects have managed to implement encrypted protocols using Chacha20 with Poly1305 and 256-bit keys with the advanced encryption standards, 
    respectively~\cite{Cipher}. It is critical to 
    standardize and enforce encryption 
    for all communications, including UAS originating or terminating data 
    to prevent eavesdropping, location tracking, data breaches, and other attacks to privacy and security integrity.    
    \item \textbf{USS/UTM A\&A:} 
    Most of the studied threat models and solutions target the UAS nodes. The 3GPP standards do not 
    provide details on the USS/UTM authentication. The USS/UTM are the main components within the UAS framework where most of the authentication, authorization, and other related information about the UAV and UAV-C are 
    stored and processed. The 3GPP 
    assumes that the USS/UTM 
    is 
    a trusted node 
    prior to the network authentication of the UAS nodes. 
    Such assumption 
    may be exploited by an adversary to perform a 
    variety of attacks, such as USS/UTM spoofing and requesting network services for unauthorized UAS missions. 
    It is important to perform an authentication check for the USS/UTM within The 3GPP to confirm its identity and parameters. 
    \item \textbf{A\&A lifetime:} The 3GPP has established revocation procedures to update the A\&A parameters or processes; however, it does not define a specific lifetime and when a revocation shall be triggered. It is triggered 
    only when a node requests it. Attackers 
    can take the advantage of potentially long-lived authentication parameters 
    and use them 
    to provide 
    access 
    to malicious nodes and perform adversary attack or flooding attacks to degrade the performance of the system. 
    The A\&A revocation process 
    should therefore be regularly triggered 
    to maintain an up-to-date status 
    of the active UAS nodes and missions.  
    \item \textbf{Blockchain for UAV security:} The standards committees should investigate the use of blockchain and distributed ledger technologies to support the registration of UAS nodes 
    with desirable characteristics such as non-repudiation and tunable tradeoffs between operator privacy and public transparency. The blockchain 
    can supplement flight data recording to ensure that the data exchange over the cellular network 
    is secure, tamper proof, and traceable for the entire UAS mission 
    without human intervention. 
\end{itemize}
\section{Conclusions}
\label{sec:conclusions}
UAVs, UAV-Cs, and the USS/UTM need to be connected reliably and securely. In order to enable flexible and safe operations of UAVs, the cellular communications network is being considered to carry UAS data and control signals and the corresponding interfaces and protocols are being standardized for emerging 5G networks. This paper has 
presented the 3GPP 
architecture for UASs connected to the 5G network and has discussed the critical security threats, the 3GPP procedures, and the remaining research and standardization opportunities related to A\&A, location information privacy, and C2 signaling. 
Research needs to feed 
the standardization process of this rapidly evolving technology. 
Experimental research can further highlight the importance of rigorously specifying the 
security framework procedures, parameters, and configurations in the standards and ensuring that they are fully implemented and tested. 

\section*{Acknowledgement}
This work was supported in part by the National Science Foundation under grant number CNS-1939334. 


\balance

\bibliographystyle{IEEEtran}
\bibliography{Refs,vuk}
\section*{Biographies}
\small
\noindent
\textbf{Aly Sabri Abdalla} (asa298@msstate.edu)
is a PhD candidate in the Department of Electrical and Computer Engineering at Mississippi State University, Starkville, MS, USA. His research interests are on scheduling, congestion control and wireless security for vehicular ad-hoc and UAV networks.

\vspace{0.2cm}
\noindent
\textbf{Vuk Marojevic} (vuk.marojevic@msstate.edu) is an associate professor in electrical and computer engineering at Mississippi State University, Starkville, MS, USA. His research interests include resource management, vehicle-to-everything communications and wireless security with application to cellular communications, mission-critical networks, and unmanned aircraft systems.
\end{document}